\documentclass[prc,reprint,showpacs,preprintnumbers,amsmath,amssymb,superscriptaddress]{revtex4-1} % normally use preprint. but the use of reprint gives a journal view
\usepackage{graphicx}
\usepackage{dcolumn}
\usepackage{bm}
\usepackage{floatrow}
\usepackage[caption=false]{subfig}
\usepackage{booktabs}
\usepackage{multirow}
\usepackage{siunitx}
\usepackage{subfig}
\usepackage{longtable}
\usepackage{footnote}
\usepackage{afterpage}
\usepackage{adjustbox}

\begin{document}
\preprint{}

\title{Evolution of the N=20 and 28 Shell Gaps and 2-particle-2-hole states in
the FSU Interaction. }
 
\author{R. S. Lubna}
\altaffiliation{Present  address: TRIUMF, Vancouver, BC V6T 2A3, Canada.}
\affiliation{Department of Physics, Florida State University, Tallahassee, FL 32306, USA}
\author{K. Kravvaris}
\affiliation{Department of Physics, Florida State University, Tallahassee, FL 32306, USA}
\affiliation{Lawrence Livermore National Laboratory, Livermore, CA 94550, USA}
\author{S. L. Tabor} 
\affiliation{Department of Physics, Florida State University, Tallahassee, FL 32306, USA}
\author{Vandana Tripathi}
\affiliation{Department of Physics, Florida State University, Tallahassee, FL 32306, USA}
\author{E. Rubino}
\affiliation{Department of Physics, Florida State University, Tallahassee, FL 32306, USA}
\author{A. Volya}
\affiliation{Department of Physics, Florida State University, Tallahassee, FL 32306, USA}

\date{\today}

\begin{abstract}
The connection between fundamental nucleon-nucleon forces and the observed many-body structure of nuclei is a main question of modern nuclear physics. Evolution of the mean field, inversion of traditional shell structures and structure of high spin states in nuclei with extreme proton to neutron ratios are at the center of numerous recent experimental investigations targeting the matrix elements of the effective nuclear Hamiltonian that is responsible for these phenomena. The FSU $spsdfp$ cross-shell interaction for the shell model was successfully fitted to a wide range of mostly intruder negative parity states of the $sd$ shell nuclei.  
In this paper we explore the evolution of nuclear structure in and around the “Island of Inversion” (IoI) where low-lying states involve cross-shell particle-hole excitations. We apply the FSU interaction to systematically trace out the relative positions of the effective single-particle energies (ESPE) of the $0f_{7/2}$ and $1p_{3/2}$ orbitals forming the $N = 20$ and $28$ shell gaps. 
We find that above a proton number of about 13  the $0f_{7/2}$ neutron orbital lies below that of $1p_{3/2}$, which is considered normal ordering, but systematically, for more exotic nuclei with lower $Z = 12$ and $10$ the order of orbitals reversed. 
The crossing of the neutron orbitals happens right near the neutron separation threshold. Our Hamiltonian reproduces remarkably well the  absolute binding energies  for a broad range of nuclei, and the inversion in the configurations of nuclei inside the IoI.  The new effective interaction accounts well for the energies and variations with mass number $A$ of  aligned high-spin states that involve nucleon pairs prompted across the shell gap. 

This work puts forward an empirically determined effective Hamiltonian where data from many recent experiments allowed us to significantly improve our  knowledge about cross shell nuclear interaction matrix elements. The quality, with which this Hamiltonian describes the two-particle two-hole (2p2h) cross shell excitations, binding energies, and the physics of aligned states that were not a part of the fit, is remarkable, making the FSU interaction an important tool for the future exploration of exotic nuclei.

\end{abstract}

\maketitle

\section{Introduction}

Recent experimental works in the $1s0d$ shell with large $\gamma$ detector arrays and heavy-ion fusion reactions have substantially extended the knowledge of relatively high spin states.  However, these do not form well-behaved rotational bands amenable to study by collective models because rotational energies are comparable to single-particle energies.  On the other hand microscopic configuration-interaction model calculations are feasible in these lighter nuclei.  The USD family of effective interactions \cite{usd, usdAB} have been very successful in describing most lower-lying positive-parity states of nuclei with $8 \leq (N, Z) \leq 20$.  However, higher spin states involve excitations into the $fp$ shell where orbitals contributing larger values of angular momentum are occupied, which is beyond the scope of the USD interaction.  Also, neutron-rich isotopes quickly move beyond the $sd$ shell boundaries \cite{MOTOBAYASHI, huber, detraz, GUILLEMAUD, YANAGISAWA, vandana32Mg}.  

Over the years, several configuration interaction models have made significant contribution towards explaining cross-shell excitations \cite{sdpf-nr, sdpf-u, psdpf, sdpf-m, sdpfumix}. A case in point is the ``Island of Inversion" (IoI).  Perhaps in an inverse way the first contribution came from the failure of the otherwise very successful pure $sd$ interactions \cite{usd, usdAB} to reproduce the stronger binding energy measured for $^{31}$Na \cite{thibault}, pointing to the importance of effects outside the $sd$ shell.  The WBMB \cite{wbmb} interaction, which was designed for the nuclei near $^{40}$Ca was successful in reproducing the inversion of some nuclei within the IoI.  More recent shell model calculations using interactions like SDPF-M \cite{sdpf-m}, SDPF-U-MIX \cite{sdpfumix} have shown that the IoI phenomenon can be accounted for by a reduction of the $N = 20$ shell gap. Recently, a significant theoretical result was reported, see  Ref. \cite{tsunoda}, showing the emergence of IoI effect from nucleon-nucleon forces stemming from the fundamental principles of QCD.  This highlights the importance of certain cross $sd$ - $fp$ interaction terms that we assess in this work using experimental systematics. 

In search for a single cross-shell interaction which works well over a wide range of nuclei, we have developed a new interaction \cite{fsu-38Cl} with parallel treatment of protons and neutrons by fitting the energies of 270 states in nuclei from $^{13}$C through $^{51}$Ti and $^{49}$V originated from the WBP interaction \cite{wbp} using well-established techniques. The present report is organized as follows: First we will discuss the development of the new FSU shell model interaction. The trend of the effective single particle energies (ESPEs) of the $0f_{7/2}$ and $1p_{3/2}$ orbitals for the $sd$-shell nuclei will be examined along with the comparison to the experimental data. Then we will move to the IoI region and test some predictions of the FSU interaction in this region. Finally, the experimentally observed fully aligned states with the $f_{7/2}^2$ configuration will be interpreted with the new shell model interaction.

%%%%%%%%%%%%%%%%%%%%%%%%%%%%%%%%%%%%%%%%%%%%%%%%
\section{Development of the FSU empirical shell model interaction}

A modified version of the WBP \cite{wbp} interaction has been used as the starting point of the data fitting procedure. The WBP interaction was developed in order to address the cross-shell structure around $A = 20$. 
While the $sd$-$fp$ cross-shell matrix elements of the WBP were taken from the WBMB \cite{wbmb} interaction which was developed for the nuclei around $^{40}$Ca, the different single particle energies and different implementation make WBP  not good for the upper $sd$-shell nuclei.  Yet, the WBP is a perfect starting point for a more modern, much broader assessment of the nuclear matrix elements. Our data set included nuclei from the upper mass region of the $p$, the full $sd$, and lower mass region of the $fp$ shells; where we systematically looked at states that involve a particle promotion across the harmonic oscillator shell, referred to as 1 particle- 1 hole excited states (1p1h).
In the $sd$-shell region of most interest and most data, the combined 0p0h and 1p1h space considered  in the fit is 
equivalent to $0\hbar\omega$ and  $1\hbar\omega$ often referred to as the $N_{\rm max}=1$ harmonic oscillator basis truncation. 
The resulting fit seamlessly spans from the $A=20$ region, where the low-lying intruder states are predominantly those with holes in the lower $p$-shell, to the island of inversion around $A=40$ where particles are promoted to $fp$ shell. 
The ability to separate the center of mass exactly within the $N_{\rm max}=1$ harmonic oscillator basis truncation is an additional benefit of this strategy. 

Before the current effort of developing the FSU interaction, a number of attempts have been made to modify the WBP interaction, mainly by changing the single particle energies (SPEs) of the $fp$-shell orbitals for a particular $sd$-shell nucleus and applying that for the nearby isotopes. For example, in the WBP-A \cite{wbp-a} version, the SPEs of the $f_{7/2}$ and $p_{3/2}$ orbitals were lowered in order to better explain the negative parity intruder states of $^{34}$P. This adjustment was quite successful in explaining the energy levels of $^{32}$P  and $^{36}$P, however, WBP-A failed to predict the intruder states of $^{31}$Si. Hence, another version of the WBP, called the WBP-B was introduced \cite{wbp-b} by changing the SPEs of the $f_{7/2}$, $p_{3/2} $ and $p_{1/2}$ orbitals.  In a different version, named WBP-M \cite{wbp-m}, all the SPEs of the $fp$ shell orbitals were changed in order to reproduce the energies and the ordering of the $3/2^-$ and $7/2^-$ states of $^{27}$Ne which eventually fixed the ordering of the same levels in $^{25}$Ne and $^{29}$Mg. 
However, none of these modified versions was able to reproduce the experimental data for a large range of the nearby nuclei, and hence we have taken a step forward towards building a more general effective shell model Hamiltonian.

The model space for the WBP interaction and for the newly developed one consists of four major oscillator shells; $0s$, $0p$, $1s0d$, and $0f1p$.
The following steps briefly describe the development of the FSU interaction.
\begin{itemize}
\item The newly developed interaction starts from the WBP framework, the model consists of four major oscillator shells: $0s$, $0p$, $1s0d$, and $0f1p$. 
Isospin invariance is assumed and Coulomb corrections to the binding energies are implemented using the standard procedures as discussed in Refs. \cite{wbmb,usd,usdAB}.
\item The single particle energies (SPE) and the two body matrix elements (TBME)  of the $0s$ and $0p$  shells and across
$0s$ - $0p$ are same as those of the original WBP interaction and are not a part of the fit. 
\item The TBMEs within the $sd$ shell are taken from the USDB \cite{usdAB} interaction and also are not a part of the fit. 
\item The 6 monopoles between the orbitals of the $0p$ shell and $sd$ shell are modified simultaneously with the $sd$ shell single particle energies, thus changing the shell gap but ensuring that excitation energies of all $0\hbar \omega$ states in the $sd$ shell are identical to those from the USDB calculations. 
\item $sd$ - $fp$ cross-shell matrix elements:
\begin{enumerate}
\item $1p_{1/2}$ orbital in the $fp$ shell is relatively high and not very sensitive to our data set. We thus fitted only one monopole term  between the $1p_{1/2}$ and the $sd$ orbitals. This amounts to two fit parameters because we have allowed different strengths for isospin T=0 and T=1. 
\item Only the monopole terms between $0f_{7/2}$ - $0d_{5/2}$ and $1p_{3/2}$ - $0d_{5/2}$ were considered since $d_{5/2}$ is deeply bound for $sd$ - $fp$ cross-shell nuclei. A total of 4 parameters were varied for T=0 and T=1. 
\item For the remaining $0f_{7/2}$ - $1s_{1/2}$, $0f_{7/2}$ - $0d_{3/2}$, $1p_{3/2}$ - $0s_{1/2}$, and $1p_{3/2}$ - $0d_{3/2}$ all multipole-multipole density terms were fitted. A total of 24 parameters were varied. 
\end{enumerate}
\item For the $fp$ shell, GXPF1A \cite{gxpf1a} was used as a starting Hamiltonian and all the TBMEs associated with only $0f_{7/2}$ and $1p_{3/2}$ were fitted; a total 30 TBMEs and hence 30 parameters were adjusted within the $fp$ shell.
\item All the matrix elements within the $sd$ and $fp$ shells as well as the $sd$ - $fp$ cross shell were scaled with $A^{-0.3}$. However, no scaling was adopted for the cross shell interactions between the lower $p$ and the $sd$ shells.
\item A total of 70 parameters were fitted using 270 experimentally observed states compiled in Ref. \cite{lubnaThesis} and \cite{nndc}. The experimental data was compiled from four groups 

\begin{enumerate}
\item Intruder states sensitive to $p$ - $sd$ shell gap. This group consists of pure $p$ shell C and N isotopes and 
nuclei between O to Si with states that have strong spectroscopic factors if populated via $(p,\, d)$ reactions.
\item Negative parity states in $sd$-shell populated via $(d,\, p)$ reactions which are sensitive to particle promotion from $sd$ to $fp$. High spin states, that gain spin from  the promotion of a particle to $0f_{7/2}$  are of particular importance. 
\item Neutron rich cross shell nuclei with $Z<20$ and $N>20$ where both $0\hbar\omega$ and $1\hbar\omega$ types of states were included in the fit. 
\item Nuclei in $fp$ shell with $Z\ge 20$ and $N\ge 21$; the $0\hbar\omega$ states in these nuclei are critical for tuning the $0f_{7/2}$ - $1p_{3/2}$ gap. 
\end{enumerate}

\item The fitting procedure followed the method described in [2], with 40 linear combinations of parameters being selected at each iteration. We reached the convergence after 6 iterations with an overall rms deviation from experiment of 190 keV. 
\item All calculations were carried out within $N_{\rm max}=1$ truncation thus including $0\hbar\omega$ and  $1\hbar\omega$ types of excitations that due to different parities do not mix. This truncation allows for exact identification and separation of the spurious center-of-mass excitations.  
\item Tables of the matrix elements can be found in the Thesis publication of  Lubna, Rebeka Sultana \cite{lubnaThesis}. Users are encouraged to contact the authors for help with the calculations, further details and updates. 
\end{itemize}
 
\begin{figure}[h!]
\begin{center}
\includegraphics[scale=0.3]{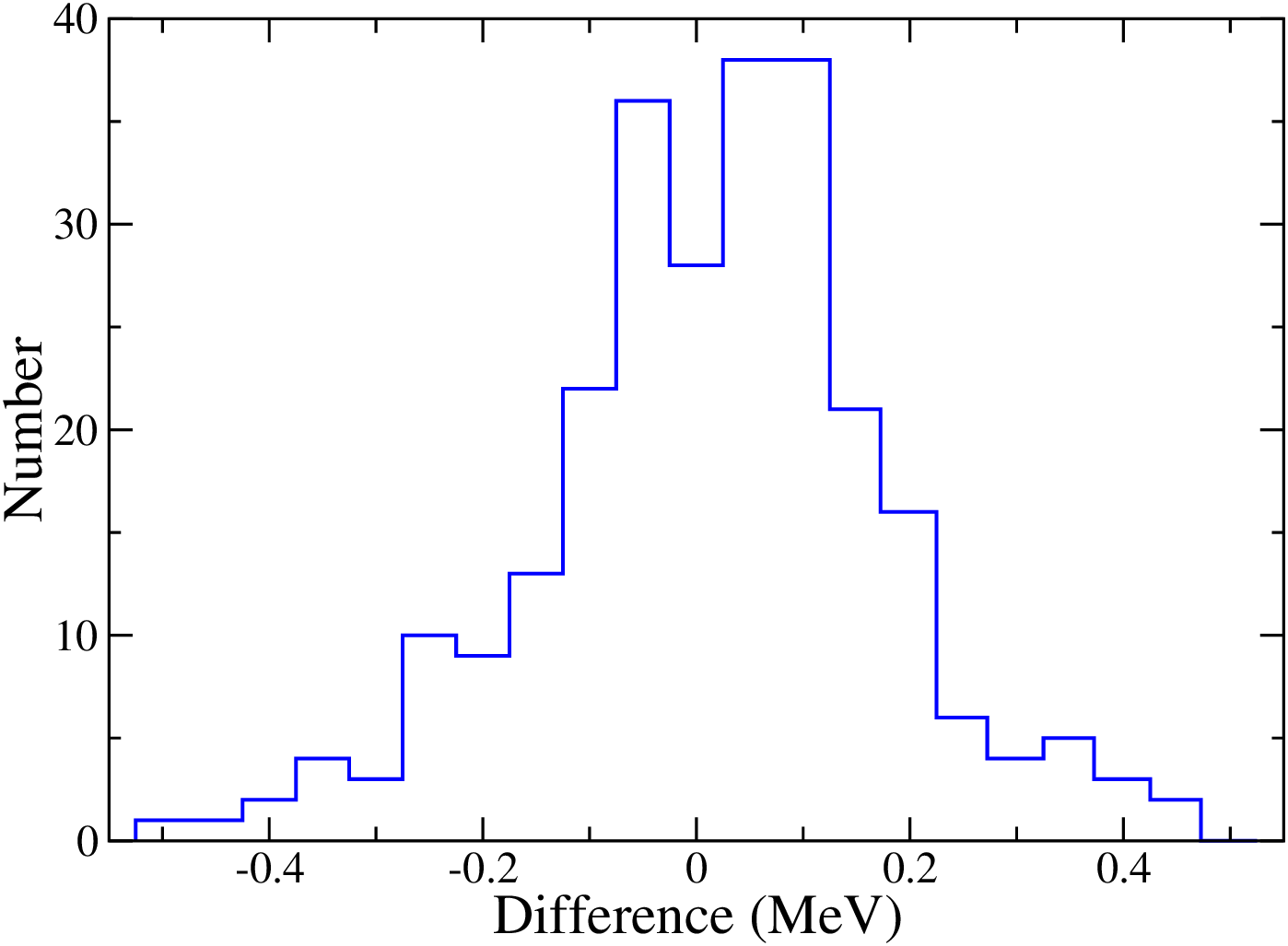}
\end{center}
\caption{Histogram of the differences in excitation energy between experiment and the FSU interaction fit.  The root-mean-square
deviation is 190 keV.}

\label{fig:fsu9Differences}
\end{figure}

 All the shell model calculations were performed with the shell model code CoSMo \cite{COSMO}. A histogram of the differences between the experimental states included in the fit and those predicted with the FSU interaction is shown in Figure \ref {fig:fsu9Differences}.

%%%%%%%%%%%%%%%%%%%%%%%%%%%%%%%%%%%%%%%%%%%%

\section{Effective Single Particle Energy (ESPE)}

The evolution of the mean field, which is described by the position of the single particle levels and how they change
with number of protons and neutrons, is a particularly interesting and non-trivial question in the strongly-interacting two-component many-body systems of atomic nuclei. In most nuclei the single-particle strength is distributed over many states. 
Systematic studies have been performed before with other shell model interactions \cite{sdpf-m, sdpf-u, smirnova} to understand the evolution of the ESPE. An experimental approach of determining the ESPEs has been to measure and sum up the energies of appropriate states weighted by the reaction spectroscopic factors.  
This process is limited by decreasing cross sections for higher lying states and by difficulties in making spin assignments and in determining what fraction of the cross sections come from direct reaction components. 

Theoretical approaches do not suffer from most of these experimental limitations, but have their own uncertainties.  Perhaps chief among them being the uncertainty in the interaction Hamiltonian.  The bare single-particle energies tell only a part of the story of the effective shell positions.  The TBME  have a major influence on the positions of the orbitals.  In fact, the TBMEs shift the orbitals based on the number of particles in shells, and are the major reason that one interaction could fit such a wide range of nuclei. 

How the newly developed FSU interaction describes the shell evolution is among the most interesting immediate questions that can be addressed. While the FSU interaction was fitted to the negative-party states in $sd$ nuclei, 
the study of the ESPE extrapolates to a much  broad spectrum of configurations not limited by those experimentally reachable with single-nucleon transfer reactions. 

The evaluation of the ESPE relies on $0\hbar\omega$ and  $1\hbar\omega$ calculations. 
In order to determine the ESPE of the $0f_{7/2}$ and $1p_{3/2}$ orbitals, we have followed a procedure similar to the experimental approach, but using the theoretically computed energies and spectroscopic factors 
in the following formula

\begin{equation}
\label{eq:eqn1}
\rm{ESPE}=\frac{\sum_{i=1} {\rm SF}_i \times E^*_i}{\sum_{i=1}{\rm SF}_i}
\end{equation}
In the above formula, $\rm {SF}_i$ is the spectroscopic factor for
$A\rightarrow A+1$ where a particle is placed onto a single-particle orbit of interest
above an even-even $A$ core. The  $\rm{E}^*_i$ is the excitation energy of the i-th state in $A+1$ with the matching quantum numbers  measured relative to the ground state energy of the core $A.$ 
It has been observed from the calculations that it is enough to consider 30 lowest states in the sum (\ref{eq:eqn1}), 
by then the SF reach to a saturation and the ESPE converges. 
From the formal theoretical perspective, Eq. (\ref{eq:eqn1}) represents single particle energies of the mean field  arising from the shell model Hamiltonian. 

\begin{figure}[h!]
\begin{center}
\includegraphics[scale=0.33]{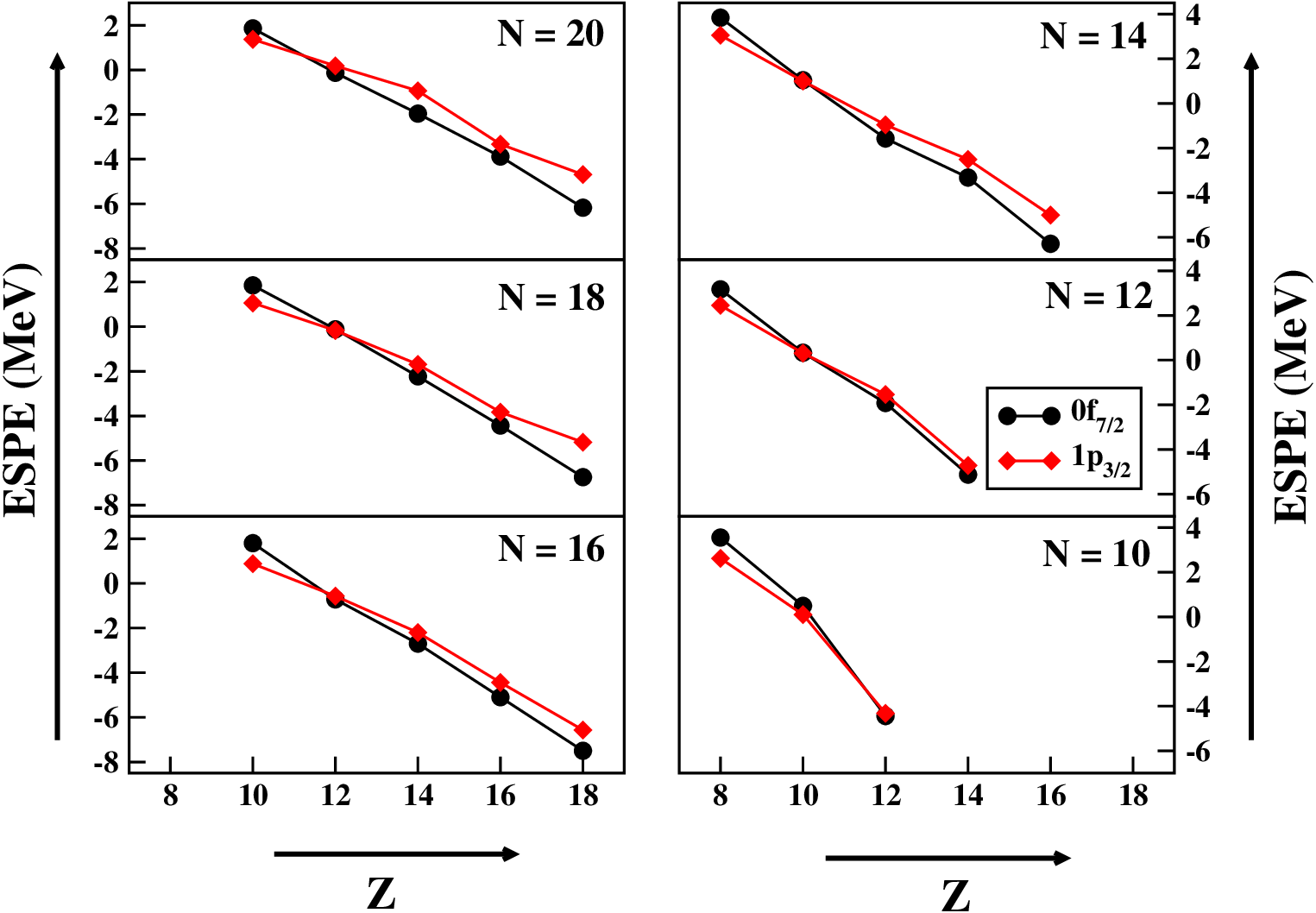}
\end{center}
\caption{Neutron Effective Single Particle Energies (ESPEs) of $0f_{7/2}$ and $1p_{3/2}$ orbitals calculated with the FSU interaction. They represent the theoretical centroids of the energies of the $0f_{7/2}$ and $1p_{3/2}$ orbitals.  In the ``normal" ordering the red diamonds ($1p_{3/2}$) lie above the black circles ($0f_{7/2}$). }
\label{fig:ESPE}
\end{figure}

The ESPEs obtained from the above formula across the $sd$ shell are plotted in Figure \ref{fig:ESPE} as a function of proton number $Z$.  The points represent the ESPEs of the $0f_{7/2}$ and $1p_{3/2}$. The systematic crossing of the ESPEs of the $0f_{7/2}$ and $1p_{3/2}$ orbitals with increasing neutron number is evident in the figure. The crossing occurs between $Z = 10$ and $12$, suggesting that the $N = 28$ shell gap shifts to $N = 24$ with lower $Z$, which points to the inversion of $0f_{7/2}$ and $1p_{3/2}$ neutron orbitals. The ground state of $^{31}$Ne is tentatively assigned $3/2^-$ as is the first excited state in $^{27}$Ne \cite{nndc}. In $^{27}$Mg the lowest $3/2^-$ and $7/2^-$ states are essentially degenerate \cite {nndc}.\\ \\

This inversion of the $1p_{3/2}$ and $0f_{7/2}$ ESPE is related to the 2-body interactions between nucleons in the $sd$ and $fp$ shells;  the effect of this interaction is density dependent and varies as a function of the shell fillings.  In the FSU interaction these TBME emerge as a consequence of fitting the energies of the states in a wide range of nuclei. Over half a century ago Talmi and Unna \cite{talmi} attributed the inversion of the $1s_{1/2}$ and $0p_{1/2}$ orbitals to the same principle.  Alternate explanations, especially for the $1s_{1/2}$ and $0p_{1/2}$ case, have been given in terms of the effects of weak binding on the mean field of low $\ell$ orbitals.  Hoffman ${\it et\, al.}$ \cite{hoff} have explored the weak binding effect for pure single-particle shells in a Woods-Saxon potential and have shown that it is large near the threshold for neutron $s$ states. 
While much smaller for $p$ states, there is still a crossing between the $0p_{1/2}$ and $0d_{5/2}$ orbitals at the threshold. 
 A similar effect for the $1p_{3/2}$ and $0f_{7/2}$ appears to be a contributing factor to the inversion shown in Figure \ref{fig:ESPE}. Nearly all crossings occur around ESPE$=0$ indicating that the levels become unbound. Indeed,
the centrifugal barrier for $\ell=3$ $f$ orbital is high which would make a transition into the continuum smooth, while for the $\ell=1$ $p$-wave the interaction with the continuum is strong and is pushing the level down as discussed in Ref. \cite{volyapc}. 
It appears that the continuum effect is incorporated through the fitting of the effective interaction,  but this can be a challenge for theoretical methods that do not take continuum of reaction states into account. This inversion of the $1p_{3/2}$ and $0f_{7/2}$ ESPE at high neutron excess also has implications for the IoI phenomenon discussed in the next section.

Another way of examining the systematics of shell evolution, which is closer to experiment, is from the positions of the states carrying the largest part of the single-particle strength.  Such a comparison is shown in Table \ref{tab:dp} which lists the experimental and theoretical excitation energies of the lowest $3/2^+$, $7/2^-$, and $3/2^-$ states, of the even $Z$ odd mass nuclei, along with the predicted and measured $(d, \, p)$ reaction spectroscopic factors (SF).  As mentioned before, there is more uncertainty in measuring the values of SF than excitation energies and in some cases the SF cannot (lack for appropriate targets) or have not been measured.  With this in mind, the agreement between experiment and predictions using the FSU interaction for both excitation energies and SF is generally good.  Also, the relatively large values of the SF show that these states represent the dominant single-particle states.

%%%%%%%%%%%%%%%%%%%%%5 Table 1
%\onecolumngrid
\begin{table}[]
\centering
\setlength{\tabcolsep}{0.85em}  % controls tables cell horizontal size
\renewcommand{\arraystretch}{1.28}
\caption{Comparison of the experimentally observed $7/2^-$, $3/2^-$ and $3/2^+$ states of even $Z$ odd mass $sd$-shell nuclei to the predictions by the FSU interaction. The measured spectroscopic factors were taken from the NNDC \cite{nndc}. All the experimental spectroscopic factors were compiled from the $(d, \, p)$ reactions.}
\label{tab:dp}
\begin{tabular}[c]{|c|c|c|c|c|c|}
\hline
\multirow{2}{*}{Nucleus} & \multirow{2}{*}{J$^\pi$} & \multicolumn{2}{c|}{Energy} & \multicolumn{2}{c|}{(2J+1)SF}  \\ \cline{3-6} 
 					   &						& EXP 	&Th				    & EXP		& Th  \\ \hline
\multirow{3}{*}{$^{25}$Ne}	 	& $7/2^-$ 	& 4030 		& 3957 		& 5.8		 & 4.5 \\ \cline{2-6} 
 							& $3/2^-$ 	& 3330		& 3471 		& 3.0 		& 1.9 \\ \cline{2-6}  	 
 							& $3/2^+$	& 2030		& 2044		&  1.6              & 1.8	  \\ \hline
\multirow{3}{*}{$^{27}$Ne}		& $7/2^-$ 	& 1740 		& 1634 		& 2.8 		& 3.9 \\ \cline{2-6} 
	 						& $3/2^-$ 	& 765 		& 858 		& 2.6 			& 2.4 \\ \cline{2-6}  	 
	 						& $3/2^+$	& 0			&0			& 1.7		& 2.8 	\\ \hline
\multirow{3}{*}{$^{25}$Mg} 		& $7/2^-$ 	& 3971 		& 3902 		& 2.2-3.3 	& 3.9 \\ \cline{2-6} 
 							& $3/2^-$ 	& 3413 		& 3525 		& 0.9-1.2 		& 1.5 \\ \cline{2-6} 
 							& $3/2+$		& 974		& 1098		& 0.8		& 0.9	 \\ \hline
\multirow{3}{*}{$^{27}$Mg} 		& $7/2^-$ 	& 3761 		& 3827 		& 4.6 		& 3.5 \\ \cline{2-6} 
 							& $3/2^-$ 	& 3559 		& 3644 		& 1.6 			& 2.2 \\ \cline{2-6} 	  
 							& $3/2+$		& 984		& 994		& 2.4		&1.56	\\ \hline
\multirow{3}{*}{$^{29}$Mg} 		& $7/2^-$ 	& 1430 		& 1719 		& 3.0 		& 4.4 \\ \cline{2-6} 
 							& $3/2^-$ 	& 1094 		& 1396 		& 0.4			& 2.0 \\ \cline{2-6}  	  
 							& $3/2^+$	& 0			&0			& 1.2		& 1.8	 \\ \hline
\multirow{3}{*}{$^{29}$Si} 		& $7/2^-$	& 3623 		& 3684 		& 7.0 		& 4.5 \\ \cline{2-6} 
 							& $3/2^-$ 	& 4934 		& 4373 		& 2.2 			& 2.3 \\ \cline{2-6} 
 							& $3/2^+$	& 1273		& 1285		& 3.0		& 2.7 	  \\  \hline
\multirow{3}{*}{$^{31}$Si} 		& $7/2^-$ 	& 3134 		& 2855 		& 4.8 		& 5.6 \\ \cline{2-6} 
 							& $3/2^-$ 	& 3533 		& 3435 		& 1.6			& 2.8 \\ \cline{2-6} 
 							& $3/2^+$	&0			&0			& 2.8		& 2.4 	\\  \hline
\multirow{3}{*}{$^{33}$Si}	 	& $7/2^-$ 	& 1435 		& 1452 		& 			& 6.0 \\  \cline{2-6}
 							& $3/2^-$ 	& 1981 		& 1944 		& 			& 2.9 \\  \cline{2-6}
 							& $3/2^+$	& 0			&0			& 			& 1.4	\\  \hline 							
\multirow{3}{*}{$^{35}$Si} 		& $7/2^-$ 	& 0    		& 0   		& 4.5 		& 7.4 \\ \cline{2-6} 
 							& $3/2^-$ 	& 910 		& 909 		& 2.8	 		& 3.7\\ \cline{2-6} 
 							& $3/2^+$	& 974		& 936			& 			& 	\\  \hline 				
\multirow{3}{*}{$^{33}$S} 		& $7/2^-$ 	& 2935 		& 2942 		& 4.2 		& 5.8 \\ \cline{2-6} 
 							& $3/2^-$ 	& 3221 		& 3386 		& 3.5 			& 2.3 \\\cline{2-6}
 							& $3/2^+$	& 0			&0			& 3.5		& 2.6 	  \\ \hline 
\multirow{3}{*}{$^{35}$S} 		& $7/2^-$ 	& 1991 		& 2042 		& 5.4 		& 6.4 \\  \cline{2-6} 
							& $3/2^-$ 	& 2348 		& 2409 		& 2.1 			& 2.7 \\  \cline{2-6} 
							& $3/2^+$	& 0			& 0			& 1.7		& 1.5 	 \\ \hline
\multirow{3}{*}{$^{37}$S} 		& $7/2^-$ 	& 0 			& 0 			& 5.5 		& 7.3 \\ \cline{2-6} 
 							& $3/2^-$ 	& 646 		& 573 		& 1.8 			& 3.5 \\ \cline{2-6} 
  							& $3/2^+$	& 1398		& 1303		& 			&	\\  \hline 								
\multirow{3}{*}{$^{37}$Ar}		& $7/2^-$ 	& 1611 		& 1543 		& 6.1 		& 6.3 \\ \cline{2-6} 
 							& $3/2^-$ 	& 2491 		& 2679 		& 1.8 			& 2.6 \\ \cline{2-6} 
 							& $3/2^+$	& 0			& 0			& 2.2		&1.5 	  \\ \hline
\multirow{3}{*}{$^{39}$Ar}		& $7/2^-$ 	& 0 			& 0 			& 5.0 			& 6.7 \\ \cline{2-6} 
 							& $3/2^-$ 	& 1267 		& 1186 		& 2.0 				& 2.8 \\ \cline{2-6} 
  							& $3/2^+$	& 1517		&1457		& 			& 	\\  \hline 											
\end{tabular}
\end{table}
%\twocolumngrid

\begin{figure}[h!]
\begin{center}
\includegraphics[scale=0.6]{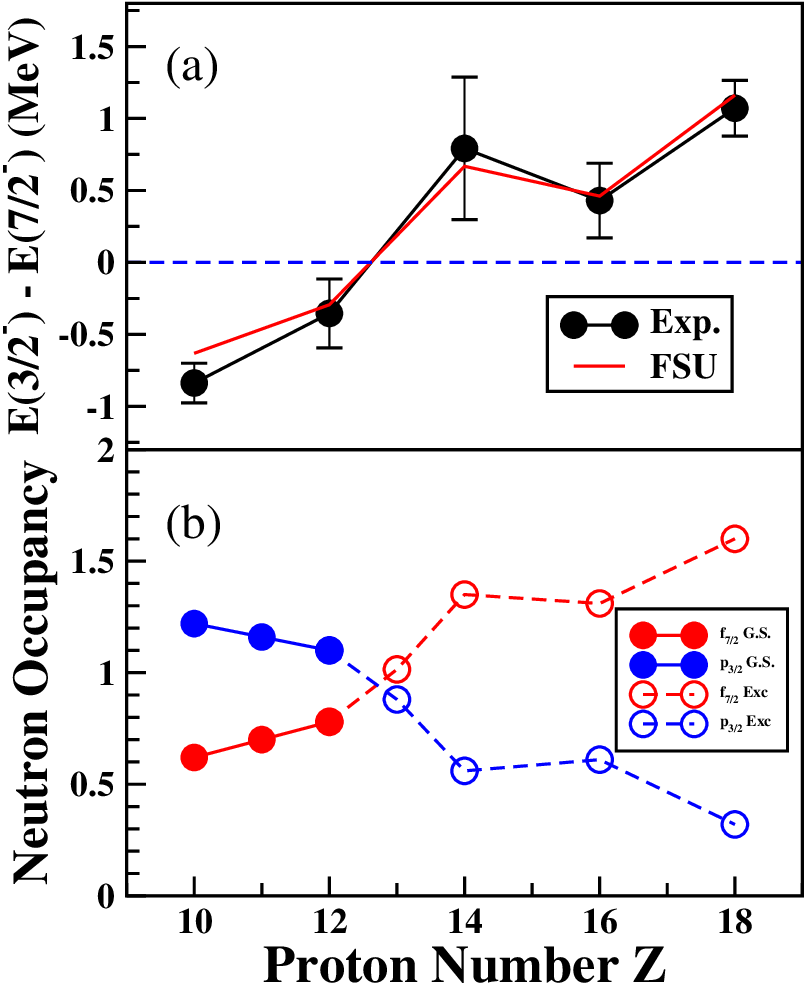}
\end{center}
\caption{(a) Average energy differences between the lowest $7/2^-$ and $3/2^-$ experimental levels in Table \ref{tab:dp}.  The error bars give an indication of the range of values for different neutron numbers.  Positive (negative) values of the ordinate correspond to the $3/2^-$ state above (below) the $7/2^-$ one.  (b) Occupancies of the neutron $0f_{7/2}$ and $1p_{3/2}$ orbitals in neutron number $N = 20$ nuclei as a function of proton number $Z$ for the lowest 2p2h states.  The values are shown as filled circles for the cases where the lowest 2p2h state is the ground state (IoI) and as open circles where the lowest 2p2h state is excited above the ground state.}
\label{fig:fpdiff}
\end{figure}

Figure \ref{fig:fpdiff}(a) provides a pictorial summary of the relative positions between the $7/2^-$ and $3/2^-$ states as a function of the proton number $Z$.  The black circles and red lines show the average values from Table \ref{tab:dp}  for experiment and theory, respectively, while the black error bars represent the variation of the experimental differences.  The observed trends are reproduced by theory, see Figure \ref{fig:fpdiff}(a).  This graph agrees qualitatively with those in Figure \ref{fig:ESPE}.  It demonstrates that the evolution of the separation between the $7/2^-$ and $3/2^-$ states is largely a function of the proton number $Z$ and that the 
$3/2^-$ energies drop below the $7/2^-$ ones between $Z=14$ and 12.  In contrast to the ESPEs which approximate the positions of the $0f_{7/2}$ and $1p_{3/2}$ orbitals the crossing between $0f_{7/2}$ and $1p_{3/2}$ happens between $Z = 10$ and 12. Together these results show that the trend is robust, but the question of the relative position of the orbitals is more complex and nuanced than was expected earlier.  

%%%%%%%%%%%%%%%%%%%%%%%%%%%%%%%%%%%%%%%%%%%%%%%
\section{Evolution of the N=20 Shell gap and the Island of Inversion (IoI)}

One of the first indications that the pure $sd$ shell model could not represent low-lying states in all $sd$ nuclei came from the experimentally measured mass of $^{31}$Na \cite {thibault}. The experimental mass was about 1.6 MeV lower than that predicted from the USD interaction \cite {usd}.  This was further clarified by the USDA, USDB showing that states in the highest $N$ - $Z$ nuclei can not be fitted.  A consistent over-prediction of 1 to 2 MeV of the ground state energies of these nuclei can be seen in Figure 9 of Ref. \cite {usdAB}.  This region of nuclei is now known as the ``Island of Inversion" (IoI) and its origin has been discussed a lot.  Most explanations center around the filled or almost filled neutron $sd$ shell and $fp$ intruder configurations leading, counter-intuitively, to lowering the energy of 
the 2 particle- 2 hole (2p2h) state, with two nucleons being promoted from $sd$ to $fp$ shell, below that of the ``normal" 0p0h. Such lowering is associated with increased correlation energy or higher deformation, lowering Nilsson orbitals.  However the effect fades away with filling of the proton $sd$ shell.

While a number of shell model calculations in the past have reproduced many aspects of the IoI, as discussed in the Introduction, here we study what the FSU interaction predicts for the Iol nuclei. 
Concentrating on the IoI region, we consider the states where two nucleons are promoted from   $sd$ to $fp$
referring to them as 2p2h states. These states were not a part of the fit and for this extrapolation to be meaningful the additional 2p2h states cannot be allowed to directly mix and renormalize the previously fitted 0p0h states. 
Due to the valence space limitation the full $2\hbar\omega$ excitations from the sd space cannot be considered.
Moreover, our tests have shown that excitations from the $0s$ and $0p$ are nearly irrelevant for the validity of this discussion, 
thus we did not include those states into our definition of 2p2h excitations. It also has been verified that the inevitable center-of-mass contamination in this truncation scheme is very low. We estimate that the errors from truncation and center-of-mass contamination amount to less then 200 keV uncertainty in the energies, which is of the same order as the rms deviation in the fit.

\begin{figure}[h!]
\begin{center}
\includegraphics[scale=0.35]{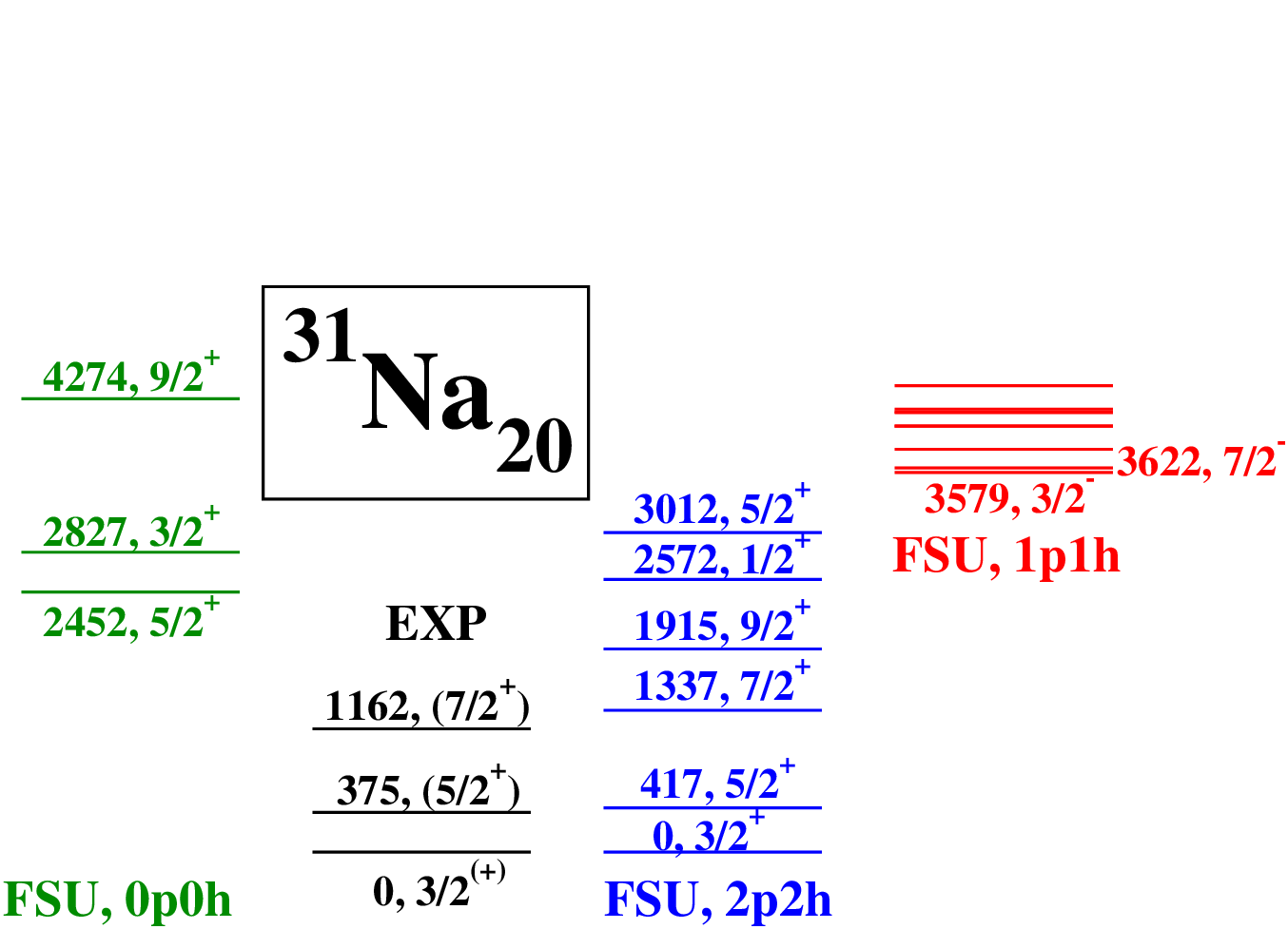}
\end{center}
\caption{The experimentally known levels of $^{31}$Na compared to the lowest ones predicted  using the FSU interaction for 0p0h, 1p1h, and 2p2h configurations.
The experimental levels agree well with the 2p2h results while the 0p0h states start almost 2.5 MeV higher in excitation energy. Only the two lowest calculated 1p1h
states are labeled because of the high level density above this.}
\label{fig:31Na}
\end{figure}

We first discuss the case of $^{31}$Na ($N=20$) \cite{thibault}. As shown in Figure \ref {fig:31Na}, the total binding energies for the first four 2p2h states were found to be below that of the lowest 0p0h state. The first three 2p2h states agree well with what is so far known experimentally, whereas 
the lowest 0p0h state ($5/2^+$) appears much higher in energy and has a different spin from the experimentally observed ground state of $^{31}$Na.

\begin{figure*} % don't use [h!] or other things if you use figure* environment
\begin{center}
\includegraphics[scale=0.55]{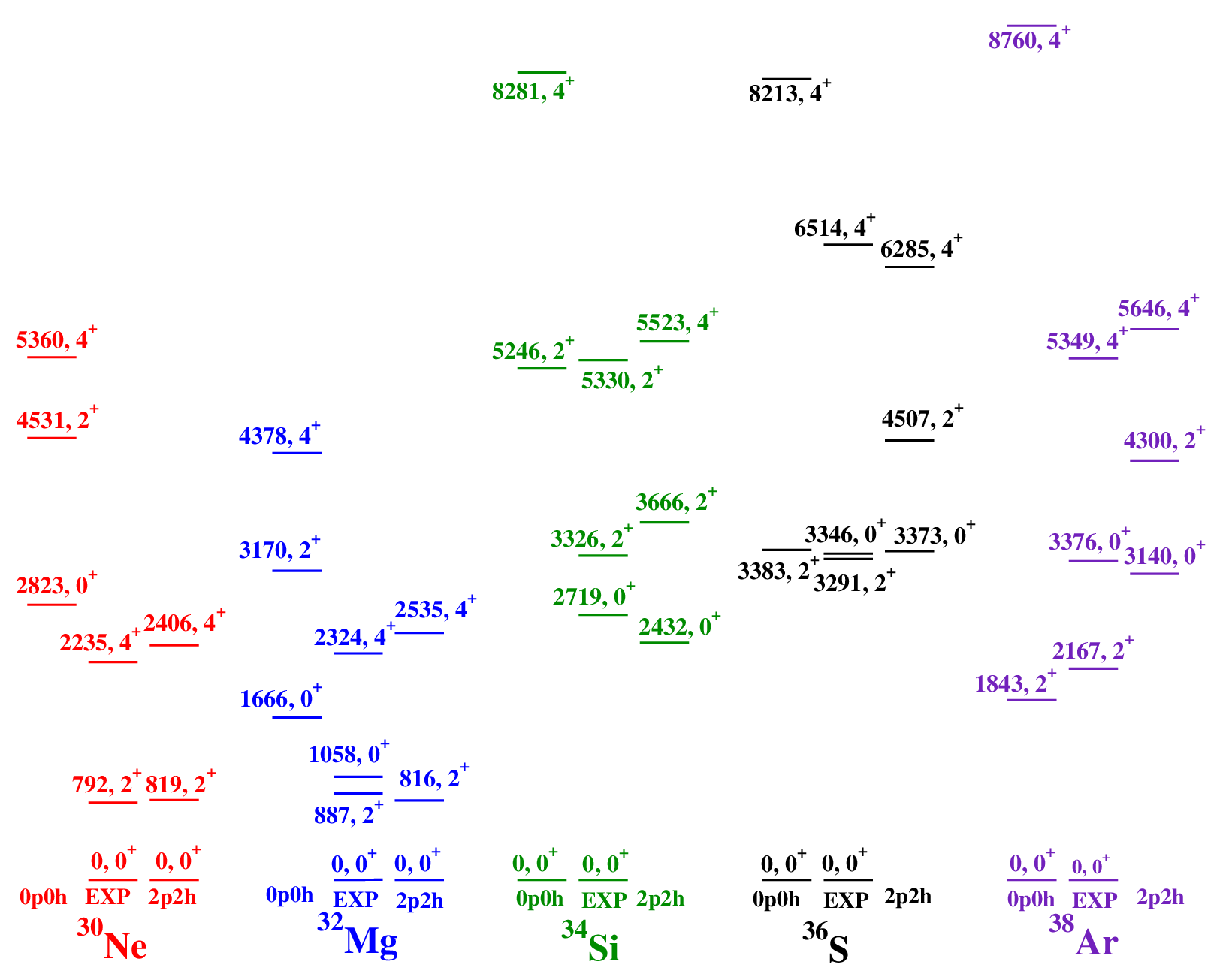}
\end{center}
\caption{The lowest experimental energy levels of $N=20$ $sd$-shell nuclei compared to those calculated using the FSU shell model interaction for 0p0h and 2p2h configurations.
The levels of the known IoI nuclei $^{30}$Ne and $^{32}$Mg agree well with the 2p2h results while the lowest states in the higher Z nuclei agree much better with
the 0p0h results.} 
\label{fig:IoI}
\end{figure*}

While the experimental information is limited, it is clear that the FSU interaction has depicted the correct picture of $^{31}$Na as one with the inverted configuration. As mentioned above, only the low $Z$ and $N \approx 20$ nuclei exhibit the IoI or inverted 2p2h - 0p0h behavior.  To explore the transition from IoI to ``normal" behavior, Figure \ref {fig:IoI} compares experimentally measured energies with our calculations for the lowest levels in a sequence of $N=20$ even-$A$ $sd$ nuclei.  For $Z = 10$ and $12$, not only do the lowest states have 2p2h character, but the whole $0^+,\, 2^+,\, 4^+$ 2p2h sequence agrees well with experiment.  In addition to starting much higher in energy, the spacing between 0p0h states differs significantly from experiment.  The story changes for $Z = 14$ $^{34}$Si, where the 0p0h $0^+$ state is substantially lower than the 2p2h one. The experimental second $0^+$ and first $2^+$ states are much closer to the 2p2h ones, while the second experimental $2^+$ level corresponds well with the 0p0h one. This shows the shape coexistence, also discussed in Ref. \cite{rotaru}.  For $Z = 16$ and $18$ both the first experimental $0^+$ and $2^+$ states correspond with the 0p0h calculations. The second $0^+$ states in both the nuclei were discussed to have 2p2h dominant configurations \cite{wood, olness, flynn}  and are in very good agreement with the FSU predictions. The $4^+$ states of $^{36}$S and $^{38}$Ar lie much closer to the calculated 2p2h ones.  Note, that the FSU cross-shell interaction describes the transition from inverted 0p0h-2p2h order to normal as a function of $Z$ despite not having been fitted to any of these states. 

This emergence of the IoI does not involve any $fp$ orbitals dropping below the $sd$ shell, at least not for spherical shape.  The lowering in energy of the 2p2h configurations does not extend so much to 1p1h ones, as shown for $^{31}$Na in Figure \ref {fig:31Na}.  The lowest 1p1h state (3579 keV, $3/2^-$) lies over an MeV above the lowest 0p0h state.  So it is the promotion of a neutron pair to the $fp$ shell which favors the 2p2h configuration so much.  The promotion of a neutron pair to the $fp$ orbital appears to lower its energy because of correlation energy in the shell model. Clearly, collective behaviors such as pairing and deformation and intricate interplay between them are central for the IoI phenomenon. Representing a mesoscopic phase transition, the picture is highly sensitive to the matrix elements of the effective Hamiltonian and in particular to the components describing short and long range limits of nucleon-nucleon in-medium interaction. 

\begin{figure}[h!]
\begin{center}
\includegraphics[scale=0.35]{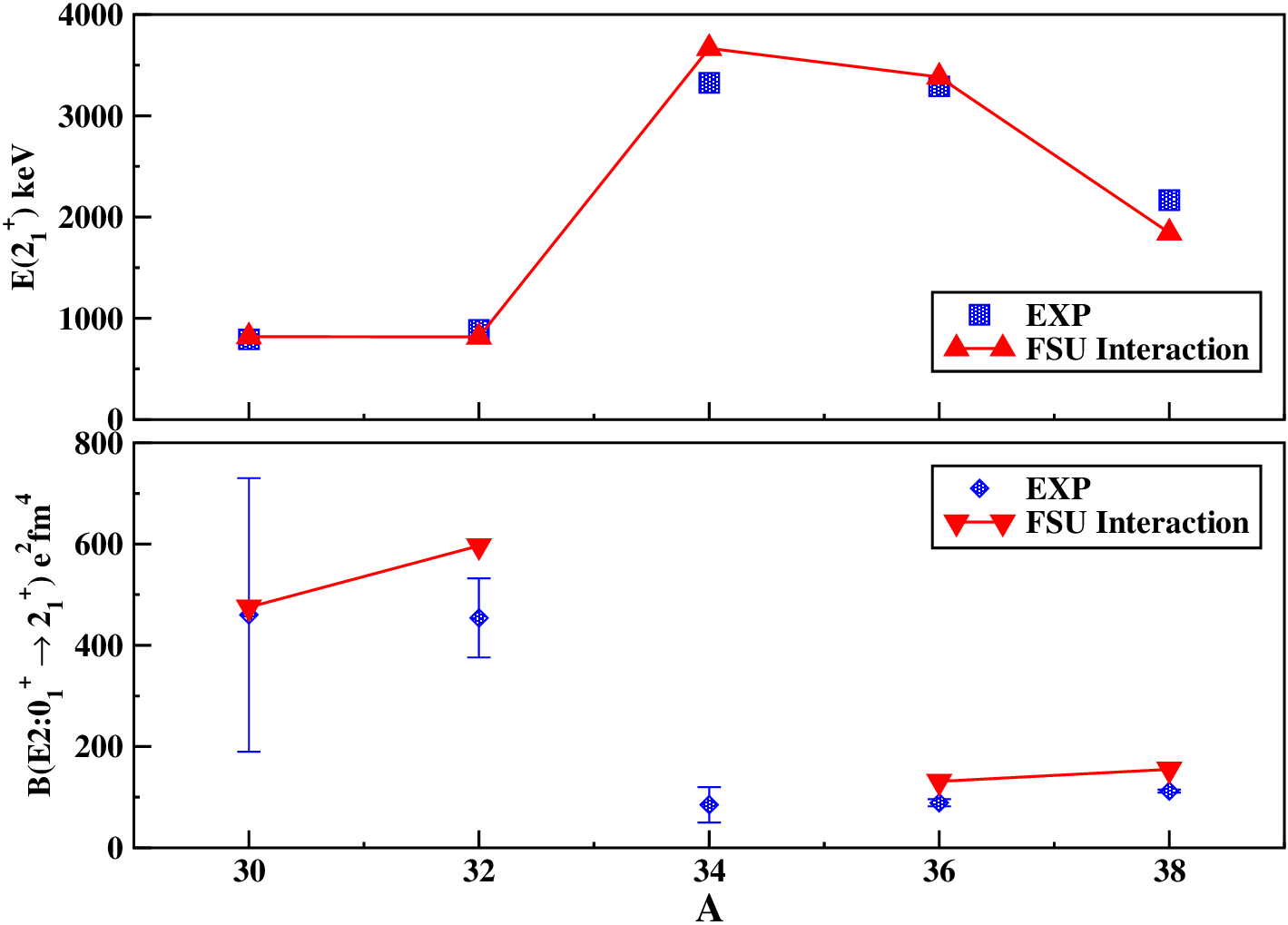}
\end{center}
\caption{Experimental E($2_1^+$) and B(E2: $0^+_1 \rightarrow 2^+_1$) values for the $N=20$ isotones are compared to those calculated by using the FSU interaction. The B(E2: $0^+_1 \rightarrow 2^+_1$)  value of $^{34}$Si has not been calculated because of the different configurations associated with the $0^+_1$ and $2^+_1$ states.}
\label{fig:BE2}
\end{figure}

In a geometrical picture IoI can be associated with increased prolate deformation due to the promotion of a pair into a down-sloping Nilsson orbital whose excitation energy decreases rapidly with increasing deformation.  An indication of this difference in deformation is shown in the lower panel of Figure \ref {fig:BE2}.  For $^{30}$Ne and $^{32}$Mg the calculated B(E2) transition strengths from the lowest $2^+$ to ground states (both of which have 2p2h configurations) are relatively large at over 400 e$^2$fm$^4$, consistent with relatively high deformation, and agree well with experiment.  In contrast the B(E2) strengths for $^{36}$S and $^{38}$Ar are rather low, consistent with near spherical shape.

\begin{figure}[h!]
\begin{center}
\includegraphics[scale=0.32]{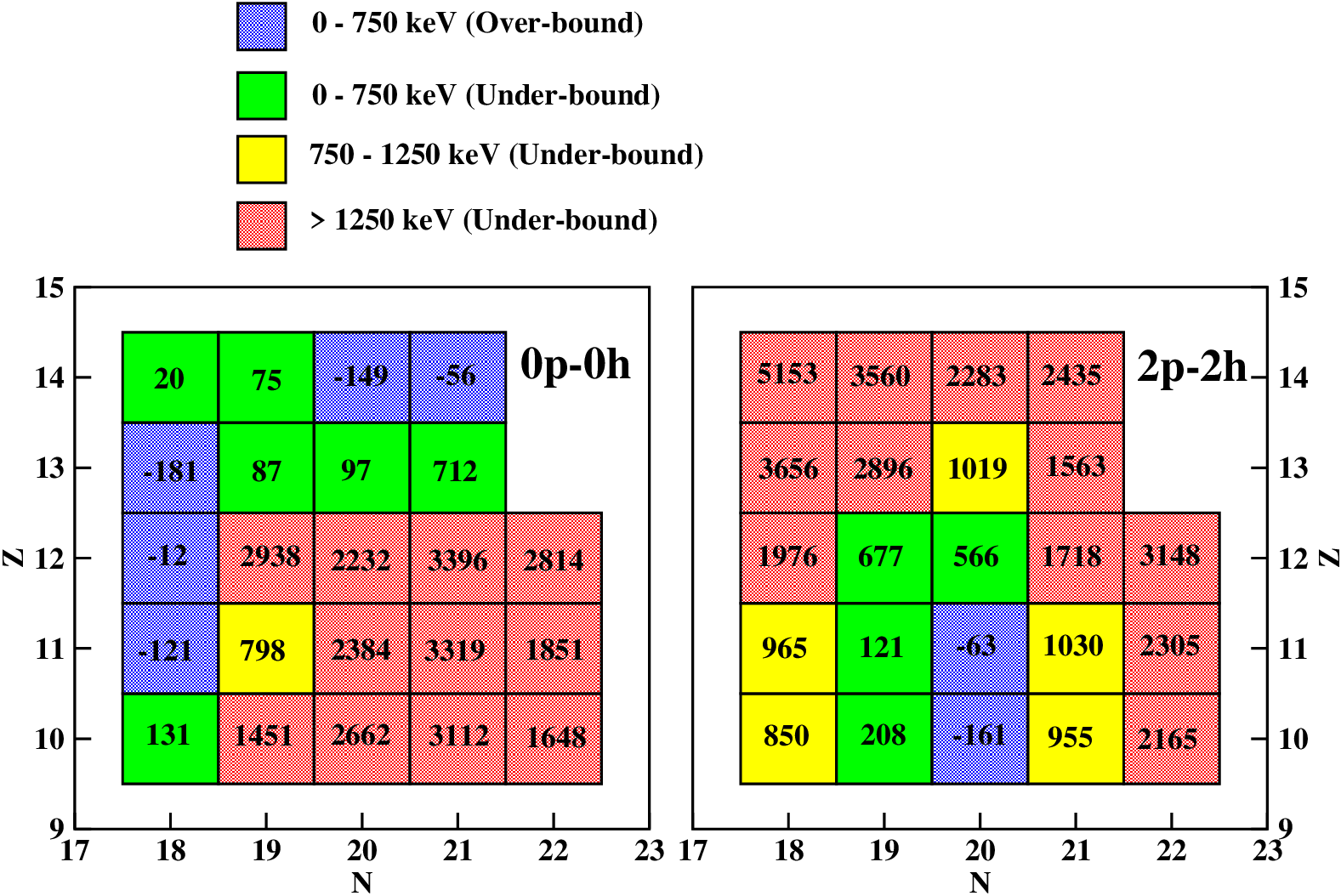}
\end{center}
\caption{The number displayed inside a box corresponding to an isotope is the difference  in binding energy between experiment and shell model predictions using the FSU interaction with 0 or 2 particle-hole configurations.  We call the states over-bound where the calculated states are more tightly bound than that of the experimental ones and under-bound when it is otherwise.}
\label{fig:BEIoI}
\end{figure}

Figure \ref{fig:BEIoI} portrays the differences between experiment and theory of the binding energies around the IoI which
are sensitive to pairing correlations. 
The calculated total binding energies are compared with the measured ground state masses from the 2016 mass evaluation \cite{mass16}. The Coulomb corrections to the total binding energies are included following procedures in Refs. \cite{wbmb,usd,usdAB}.  The $N = 21$ 0p0h (2p2h) configurations have 1(3) nucleons in $fp$, and, $N = 22$ 2p2h actually have 4 $fp$ nucleons so the $fp$ matrix elements are tested along with the cross-shell ones. Looking at the $N = 20$ isotonic chain, the agreement is quite good with an RMS deviation of 276 keV comparing the experimental binding energies with the 2p2h results below $Z = 13$ and with 0p0h for higher $Z$. For $10 \le Z \le 12$ and $19 \le N \le 21$ the 2p2h inverted configuration is lower in energy and agrees better with experiment.  Outside this range the 0p0h configuration is lower, which again agrees with experiment.  For $N = 22$ it appears that promoting a second neutron pair to $fp$ is not energetically favorable. 

A similar approach of calculating the 2p2h states was taken in Ref. \cite{wbmb} using the WBMB interaction. As mentioned earlier, the WBMB interaction was developed for the mass region near $^{40}$Ca by fitting the 1p1h states within the $sdfp$ model space. We have compared the differences in the 0p0h and 2p2h ground states calculated by using the WBMB and the FSU interactions for $N =20$ isotones in Table \ref{tab:wbmbFSU}. The predictions with the WBMB interaction were taken from Ref. \cite{wbmb}. From Table \ref{tab:wbmbFSU}, we see that both the interactions predict $^{30}$Ne, $^{31}$Na and $^{32}$Mg having their 2p2h ground state more tightly bound than that calculated for the 0p0h configurations, meaning that these nuclei are the members of the IoI. The FSU interaction predicts $^{29}$F also as an IoI nucleus, which was suggested recently by the Ref. \cite{29F}. The difference between the first two $0^+$ states in $^{34}$Si is known experimentally as 2719 keV \cite{rotaru}.  The FSU interaction predicts it better as 2432 keV. The experimentally observed $0^+_2$ states in $^{36}$S and $^{38}$Ar are at  3346 and 3376 keV respectively,  which are presumably 2p2h in nature. The FSU interaction predicts them at 3373 and 3140 keV respectively. In $^{37}$Cl the first 2p2h state was identified at 3708 keV energy \cite{37Cl}, whereas the FSU prediction is 3538 keV. The better predictability of the FSU interaction comes from a more extensive fit for a wide range of cross shell data as well as the use of a better  Hamiltonian for the $fp$ shell, we believe.  

%%%%%%%%%%%%%%%%%%%%%%%%%%%%%% Table2

\begin{table}[]
\centering
\setlength{\tabcolsep}{0.85em}  % controls tables cell horizontal size
\renewcommand{\arraystretch}{1.5}
\caption{The ground state energies with 2p2h configurations are calculated with respect to those with the 0p0h configurations using the WBMB \cite{wbmb} and the FSU interactions. The symbols W, F and T in the WBMB calculations stand for weak coupling, full WBMB space and the truncated space respectively.}
\label{tab:wbmbFSU}
\begin{tabular}{|c|c|c|}
\hline
Nucleus 		& WBMB	\footnote{Ref. \cite{wbmb}}	 & FSU  \\ \hline
\multirow{2}{*}{$^{28}$O} 	& 3038: W	& -755	\\
						& 2956: F		&	\\ \hline
\multirow{2}{*}{$^{29}$F}  	& 1286: W	& -2201	\\
						& 1338: F		&	\\ \hline
\multirow{2}{*}{$^{30}$Ne} 	& -698: W	& -2823	\\
						& -788: F		&	\\ \hline
\multirow{2}{*}{$^{31}$Na} 	& -502: W	& -2452	\\
						& -764: T		&	\\ \hline
\multirow{2}{*}{$^{32}$Mg} 	& -926: W	& -1666	\\
						& -966: T		&	\\ \hline
$^{33}$Al 				& 854: W		& 922	\\ \hline
\multirow{2}{*}{$^{34}$Si} 	& 1816: W	& 2432	\\
						& 1554: T		&	\\ \hline
$^{35}$P 				& 2698: W	& 3264	\\ \hline			
\multirow{2}{*}{$^{36}$S}  	& 3146: W	& 3373	\\
						& 3009: T		&	\\ \hline	
\multirow{2}{*}{$^{37}$Cl}	& 3195: W	&  3538 \\
						& 3091: T		&	\\ \hline	
$^{38}$Ar 				& 2701: F 	& 3140	\\ \hline							
\end{tabular}
\end{table}

\begin{figure}[h!]
\begin{center}
\includegraphics[scale=0.4]{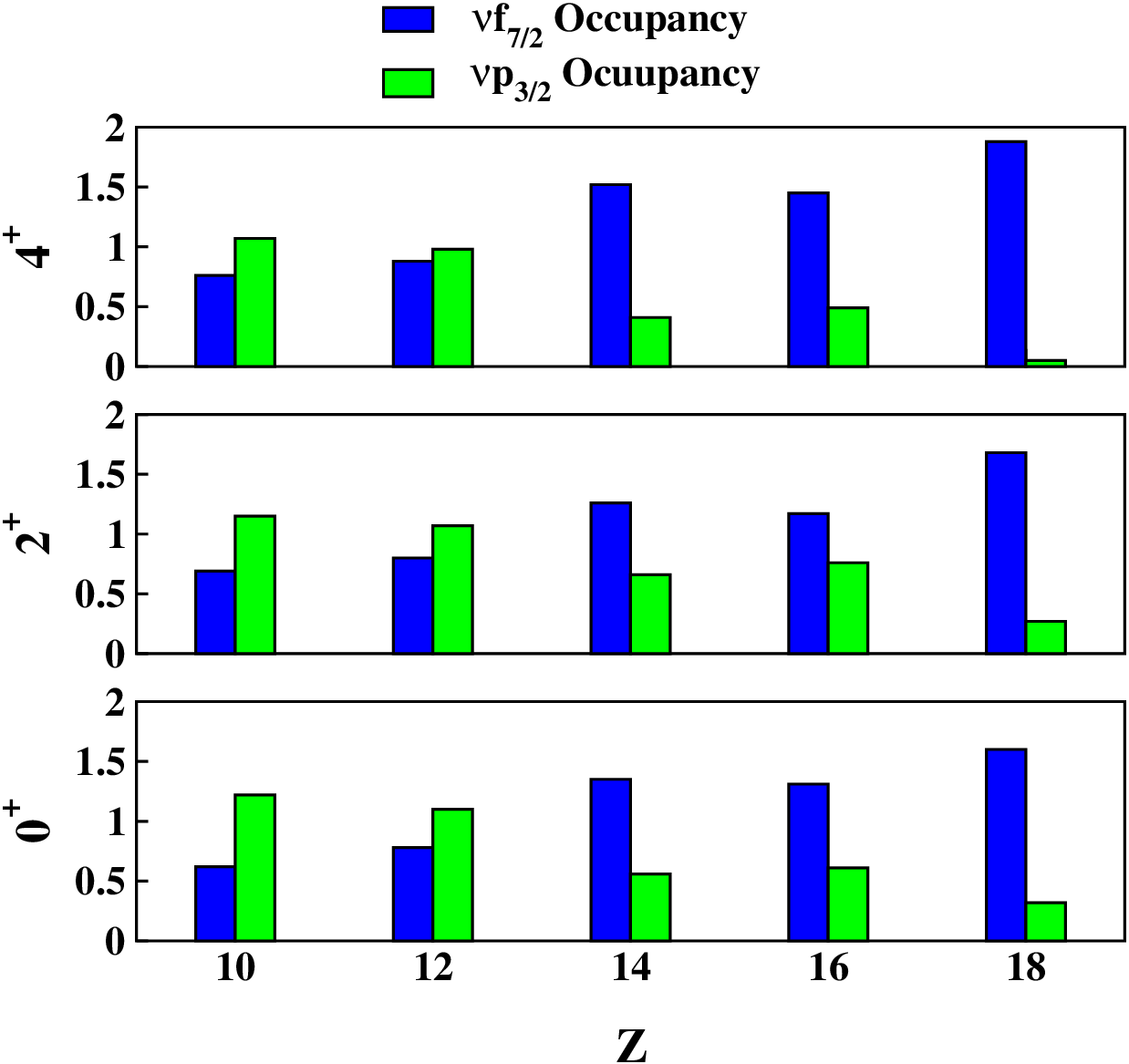}
\end{center}
\caption{2p2h occupancies of the $\nu 0f_{7/2}$ and $\nu 1p_{3/2}$ orbitals for the first $0^+$, $2^+$ and $4^+$ calculated states using the FSU interaction for nuclei with $N=20$ and $Z$ between 10 and 18.}
\label{fig:OccupIoI}
\end{figure}

Since the IoI involves excitations into the $fp$ shell, the question arises how the inversion of the $0f_{7/2}$ and $1p_{3/2}$ single particle energies at low $Z$, discussed above, affects our understanding of the IoI.  The answer, within the context of the FSU interaction is shown in Figure \ref{fig:OccupIoI}. 
This figure shows some of the $fp$ shell occupancies calculated for the lowest 2p2h states in Figure \ref {fig:IoI}. Occupancy here is defined as the average number of nucleons in a given orbital. 
There is almost no proton $fp$ occupancy calculated for these nuclei and there is a relatively constant $\nu 1p_{1/2}$ occupancy of about 0.1 neutron.  For $Z=10$ $^{30}$Ne, which is the most strongly inverted, the $\nu1p_{3/2}$ occupancy is about twice that of $\nu0f_{7/2}$.  
With increasing $Z$, the ratio of $\nu 1p_{3/2}$ to $\nu 0f_{7/2}$ decreases steadily from about 2 to about 0.2 across this region.  Of course, the energies of the 2p2h configurations rise above that of the 0p0h ones around $Z = 14$.  

We note that considering that the degeneracy of the $f_{7/2}$ is twice that of $p_{3/2}$, at the level crossing or in the limit of strong pairing the ratio of occupancies of $\nu 1p_{3/2}$ to $\nu 0f_{7/2}$ should be about 0.5. 
This indeed happens at around $Z=14$; however significant deviation from 0.5 suggests that pairing, or at least pair transfer between $f_{7/2}$ and $p_{3/2}$ is weak. Pair transfer and pair vibration, collective pairing condensation, interplay of paring and deformation in the IoI region, as well as the connection of these collective effects with the underlying matrix elements of the FSU Hamiltonian, all require more study and remain a challenge for the future. The occupancy trend is perhaps illustrated more clearly in Figure \ref{fig:fpdiff}(b) which shows the $\nu 1p_{3/2}$ and $\nu 0f_{7/2}$ occupancies of the lowest 2p2h states in the $N = 20$ nuclei as a function of proton number $Z$. Note, that for $^{34}$Si, the 2p2h $0^+$ state lies 2432 keV above the 0p0h ground state but the 2p2h $2^+$ level lies close in energy with the lowest experimental $2^+$ state.  
Together these calculations imply that the $\nu 1p_{3/2}$ orbital plays a larger role in the IoI phenomenon than does the $\nu 0f_{7/2}$ one.

\section {Fully aligned states}

In describing the states used in the fit of the FSU interaction, we included only 0p0h(1p1h) configurations for natural(unnatural) parity sectors.  In particular, no 2p2h configurations were used to adjust the interaction parameters.  After the fitting, two early tests were performed to explore the predictive properties of the FSU interaction for 2p2h configurations.  One was the calculation of the lowest 2p2h $7^+$ states in $^{34}$Cl and $^{36}$Cl \cite {fsu-38Cl}.  These agreed within 200 keV with the experimental states.  The other test was performed on $^{38}$Ar  \cite {abromeit}, since experimental states up to $8^+$ and $(10^+)$ are known.  Calculations using the USD family of interactions agree within 200 keV with the excitation energy of the lowest $2^+$ state of $^{38}$Ar, but over-predict the lowest experimental $4^+$ level by over 3 MeV.  With only two holes in the $sd$ shell, the maximum spin from coupling two $0d_{3/2}$ protons is $2 \hbar$.  The very high $4^+$ energy represents the cost of promoting a $0d_{5/2}$ proton to $0d_{3/2}$, but nature finds another less energetic way of achieving $4^+$.  This must be by promoting an $sd$ nucleon pair to the $fp$ shell.  A 2p2h calculation with the FSU interaction predicts the lowest $4^+$ level only 300 keV above the experimental one, and it predicts the $6+$ state 200 keV below experiment, while the predicted $8^+$ state is 100 keV above experiment as shown in Ref. \cite {abromeit}.

\begin{figure*}
\centerline{\includegraphics[scale=0.65,angle=0]{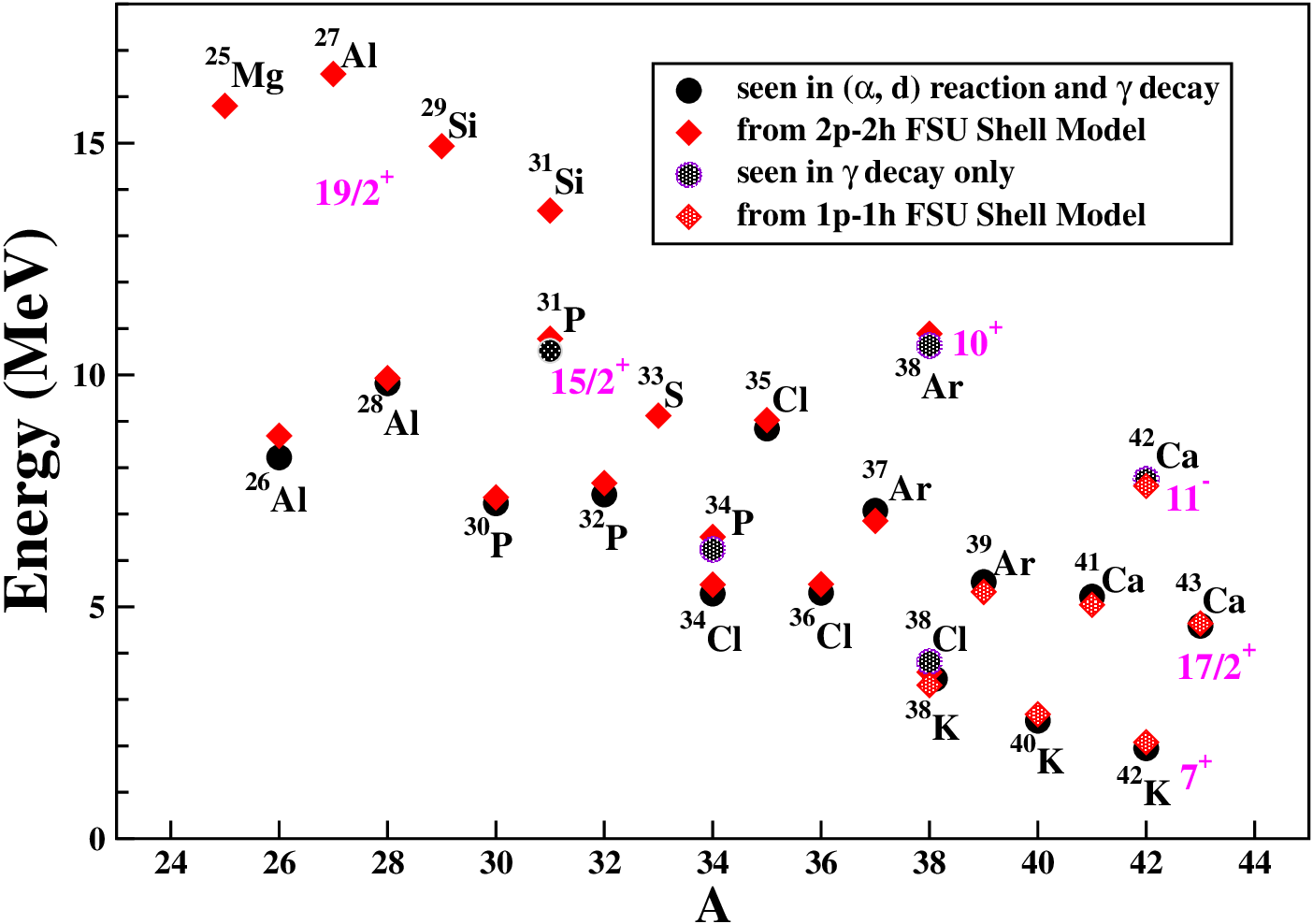}}
\caption{ Comparisons of the energies of fully aligned states in $sd$-shell nuclei with those predicted employing the  FSU interaction.  Many of the experimental points are confirmed by both selective population in $(\alpha,d)$ reactions and in high-spin $\gamma$ decay sequences and are displayed with solid black circles, while dotted black circles are used to represent states observed by only one of the two signatures.  The structure of many of these aligned states involve the promotion of two (extra) nucleons to the $0f_{7/2}$ orbital and are shown with solid red diamonds.  Those with at least one nucleon in the $0f_{7/2}$ orbital may require only one more promotion (1p1h excitation) and are shown with dotted red diamond symbols.}
\label{fig:aligned}
\end{figure*}

With this success we have searched for other states with confirmed 2p2h structure to compare with theory.  One such group of excited states across the $sd$ shell are often called the ``fully aligned" states.  One subgroup of fully-aligned states is the lowest $J^\pi = 7^+$ states.  These states have been suggested to have both odd nucleons in the highest spin orbital around - $0f_{7/2}$ - and with their spins fully aligned, which, from the Pauli principle, is only possible for non-identical nucleons.  For these calculations it is critical that the FSU interaction treats protons and neutrons on an equivalent basis.  These fully-aligned $\pi f_{7/2} \otimes \nu f_{7/2}$ are yrast and strongly populated in high-spin $\gamma$-decay sequences.  Stronger evidence of their unique nature comes from $(\alpha,d)$ reactions \cite{rivet, del, alphaD1, alphaD2, alphaD3, alphaD1962, alphaD1960}  where they are the most strongly populated states with an orbital angular momentum transfer of $\ell = 6$.  In most cases such states involve two nucleons beyond those in the dominant ground state configuration outside the $sd$ shell.  The energies of these $7^+$ states (including those in $^{34}$Cl and $^{36}$Cl mentioned above) are graphed in Figure \ref {fig:aligned} along with calculated results using the FSU interaction.  The agreement is excellent both in value and in the trend which extends from 10 MeV for the lightest nuclei down to 2 MeV for the heaviest and from 2p2h to 1p1h excitations relative to the ground state.  The calculations also indirectly confirm the spin alignment with approximately equal proton and neutron occupancies in the $0f_{7/2}$ orbitals, even though most 2p2h states in these neutron-rich nuclei as discussed in the IoI section involve predominantly two neutron configurations.

Fully aligned states are also known for some odd-$A$ nuclei where an $sd$ nucleon is also aligned in spin with the aligned $0f_{7/2}$ nucleons.  Five such cases in Figure \ref {fig:aligned} are known experimentally  as the strongest states populated in $(\alpha,d)$ reactions.  They have an unpaired nucleon in the $0d_{3/2}$ orbital which contributes an extra spin of $3/2 \hbar$.  Again the 2p2h and 1p1h calculations with the FSU interaction agree well.  In lighter odd-A nuclei the aligned $sd$ nucleon could be in the $1s_{1/2}$ or $0d_{5/2}$ orbitals, leading to total spins of 15/2 or 19/2 and higher excitation energies.  Their calculated energies are also shown in Figure \ref {fig:aligned}, but none have been seen in  $(\alpha,d)$ reactions.  A $(11/2^+,15/2^+)$ state which decays only to the lowest $13/2^+$ state and is very likely the $15/2^+$ fully aligned state has been reported \cite {nndc} in $^{31}$P, as shown in the figure would agree well with the predictions.

The last category of aligned states in the $sd$ shell consists those in even-even nuclei.  Their excitations involve the breaking of a proton and a neutron pair and promotion of one of each nucleon to the $0f_{7/2}$ orbital. For example, all 4 unpaired nucleons coupled to maximum spin of $10^+$ if both unpaired $sd$ nucleons are in the $0d_{3/2}$ orbital.  No $(\alpha,d)$ reactions to the fully aligned state in even-even nuclei are known because of the absence of stable odd-$Z$ odd-$N$ targets in the $sd$ shell.  However, the lowest experimentally known $10^+$ state in $^{38}$Ar observed by other reactions does compare well with a 2p2h calculation using the FSU interaction, as shown in Figure \ref{fig:aligned}.  In the case of $^{42}$Ca the analogous state would involve breaking a $\pi d_{3/2}$ pair, promoting one proton to $0f_{7/2}$, breaking the $\nu f_{7/2}$ pair and coupling them to maximum spin for a total of $11^-$.  This state has been seen in $\gamma$ decay following fusion-evaporation and its energy agrees well with the FSU calculation.  We hope that future experiments in the FRIB age will be able to test these predictions.  This study of aligned states targets cross shell matrix elements of high angular momentum channels that describe 
long-range effective in-medium nucleon nucleon interactions and play key role in determining nuclear shape and deformation.

\section{summary}
In this work we present an effective nuclear interaction Hamiltonian for shell model calculations, named FSU interaction.
The interaction targets a broad range of nuclei from $p$ to $fp$ shells with a particular emphasis on exotic nuclei with extreme proton to neutron ratio and on states that involve cross shell excitations. The interaction was fitted using 
binding energies and $1\hbar\omega$ states that probe cross-shell matrix elements in nuclei from $^{13}$C through $^{51}$Ti and $^{49}$V. Additional details of the fit can be found in Refs. \cite{fsu-38Cl,lubnaThesis}. 
This report provides a comprehensive study of nuclei in the region of the Island of Inversion, namely those nuclei between 
$sd$ and $fp$ shells whose low-lying structure is dominated by cross shell excitations. 

We use the newly obtained FSU interaction to infer information about the mean field and evolution of the 
effective single particle energies (ESPE). The  ESPEs of the $0f_{7/2}$ and $1p_{3/2}$ show the expected normal ordering,
where $0f_{7/2}$ is below $1p_{3/2}$ for $Z > 12$ and a consistent trend of a decreasing separation with decreasing $Z$ until the energy order reverses around $Z = 10$ to $12$. It is remarkable that the inversion happens near zero energy associated
with the decay threshold. The interaction with the continuum is not explicitly included but maybe captured as a part of the fit.
 While there have been many indications of inverted shell ordering in the past, these results present a more systematic picture from a model very firmly rooted in data.  Perhaps somewhat surprisingly, over the range explored here, the inversion appears to depend more on the proton number than on the neutron excess. 
The lowest $3/2^+$, $7/2^-$, and $3/2^-$ experimental states are surveyed for a complementary view of shell evolution.  These energies are compared with predictions of the FSU interaction in excitation energies and spectroscopic factors.  
They present a similar picture of the $0f_{7/2}$ - $1p_{3/2}$ shell evolution as a function of proton number.

The success of the FSU interaction in reproducing the negative parity states of the $sd$-shell nuclei with the $1\hbar\omega$ configuration suggests that improved, over those in Ref. \cite{rp}, calculations of the rp process rates 
 can be performed in the future. 
 
In this report, the FSU interaction was taken a step forward  and applied to configurations involving promotion of two nucleons 
from $sd$ to $fp$ (2p2h) in the region of IoI. In this region the nuclei are more tightly bound than predicted within  the pure $sd$ model space (0p0h).  The 2p2h configurations have lower binding energies and agree well with the measured ground state masses in the range $10 \le Z \le 12$ and $19 \le N \le 21$, while the 0p0h configurations are lower in energy and agree better with the measured masses elsewhere.  The lowest $2^+$ states agree well with the 2p2h calculations in the region $Z = 14$ and with 0p0h for $Z = 16 - 18$.  The results of the FSU interaction which was not fitted to these states reproduce 
well both the IoI and the transition to normal behavior. $^{34}$Si with  $Z = 14$ emerges as transitional with a 0p0h ground state and a 2p2h lowest $2^+$ state.  It would be interesting to locate the experimentally $4^+_1$ state which is predicted as 2p2h at 5523 keV.  Another implication of the FSU shell model calculations is that $\nu1p_{3/2}$ pairs dominate over $\nu0f_{7/2}$ ones in the IoI, but $\nu0f_{7/2}$ pairs dominate the lowest 2p2h states beyond the IoI. 
This is an indication of a relative weakness of pairing that would act to equilibrate occupancy. 
 Interestingly the IoI coincides relatively well with the region where the $\nu1p_{3/2}$ orbital falls below the $\nu0f_{7/2}$ one.

Another success of the FSU interaction has been the calculation of the energies and occupancies of the fully aligned states, first identified in the early 1960's in $(\alpha,d)$ reactions and frequently observed in high-spin $\gamma$-decay cascades (most involve 2p2h excitations relative to the ground state).  Their energies are reproduced very well across the mass range, and their occupancies prove the excitation of both protons and neutrons, even though pure neutron excitations are more common in other states. This is an important result that establishes values for the specific cross shell high angular momentum matrix elements that are responsible for long range effective nucleon-nucleon interaction and are particularly challenging to obtain from fundamental principles. 

This work brings forward an interesting comparison between traditional shell model interactions with those arising from first principles methods. While the former are obtained from simply fitting SPEs and TBMEs to experimental data, the latter require renormalizations, many-body forces and explicit inclusion of the reaction continuum to achieve agreement with experiment. This dichotomy, presents a modern challenge to nuclear theory and deserves a full investigation.

The capability of the FSU interaction to explain the exotic phenomena of the nuclei carries the prospect that the interaction will be successful for more exotic nuclei or states.  It is hoped that the interaction will prove valuable in the coming FRIB age. \\

\begin{acknowledgements}
This work was supported by U.S. National Science Foundation under grant No. PHY-1712953 (FSU), U.S. Department of Energy, office of Science, under Award No. DE-SC-0009883 (FSU). Part of the manuscript was prepared at LLNL under Contract DE-AC52-07NA27344.
\end{acknowledgements}

%%%%%%%%%%%%%%%%%%%%%%%%%%%%%%%%%%%%%%%%%%%%%%%%

%
\clearpage
%%%%%%%%%%%%%%%%%%%%%%%%%%%%%%%%%%%%%%%%%%%%%%%%

\end{document}